\documentclass[12pt]{article}
\usepackage{graphicx}
\usepackage{amsmath}
\usepackage{amssymb}
\usepackage{enumitem}
\usepackage{cite}
\usepackage{latexsym}
\begin{document}
\vskip 2cm

\def\warning#1{\begin{center}
\framebox{\parbox{0.8\columnwidth}{\large\bf #1}}
\end{center}}

\begin{center}

{\large {\bf Christ-Lee model: augmented supervariable approach}}

\vskip 2.5 cm

{\sf{ \bf R. Kumar$^1$ and A. Shukla$^2$\footnote{Present address: School of Physics and Astronomy, Sun Yat-Sen University, Zhuhai, 519082, China, Email:ashukla038@gmail.com}}}\\
\vskip .1cm
{\it $^1$Department of Physics \& Astrophysics,\\ University of Delhi, New Delhi--110007, India}\\
\vskip .1cm
{\it $^2$Indian Institute of Science Education and Research Kolkata,\\
 Mohanpur--741246, India}\\
\vskip .2cm
{\tt{E-mails: raviphynuc@gmail.com; as3756@iiserkol.ac.in}}\\
\end{center}

\vskip 2.5 cm

\noindent

\noindent
{\bf Abstract:} We derive the complete set of off-shell nilpotent and absolutely anticommuting (anti-)BRST as well as
(anti-)co-BRST symmetry transformations for the gauge-invariant Christ--Lee model by exploiting the  
celebrated (dual-)horizontality conditions together with the gauge-invariant and (anti-)co-BRST invariant restrictions 
within the framework of geometrical {\it ``augmented"} supervariable approach to BRST formalism. We show the (anti-) BRST and 
(anti-)co-BRST invariances of the Lagrangian in the context of supervariable approach.     
We also provide the geometrical origin and capture the key properties associated with the (anti-)BRST and 
(anti-)co-BRST symmetry transformations (and corresponding conserved  charges) in terms of the supervariables and 
Grassmannian translational generators. \\



\newpage

\section{Introduction}
The dynamics of a given physical system can be described in terms of the differential equations of various degrees. 
The classical Hamilton's equations, Schr{\"o}dinger equation in quantum theory and Maxwell's equations in electrodynamics are a few physically fundamental 
examples of such systems. 
In order to get the complete information about the dynamics of a system, one has to solve the equations which describe them. 
Interestingly,  the existence of symmetry further simplifies the solutions of physical system. This is because of the fact that one can describe 
the properties of a given system without solving all the equations of motion. Thus, the symmetry transformations are the key ingredients of modern 
physics \cite{yang}. It is well-known that the three out of {\it four} fundamental interactions of nature can be well described by the gauge 
theories and associated local gauge symmetries. Gauge symmetry is always generated by the first-class constraints present in a given physical 
theory \cite{dira,sund}.

Becchi--Rouet--Stora--Tyutin (BRST) formalism is one of the elegant, mathematically rich and unique ways to covariantly quantize any gauge
theory where unitarity and quantum gauge invariance are respected together \cite{brs1,brs2,brs3,brs4}. It is important to mention that 
for a given local gauge symmetry at the classical level, we have {\it two} global supersymmetric type (i.e., BRST and anti-BRST) 
symmetries at the quantum level \cite{cf,oji}. 
These symmetry transformations have two innate properties: nilpotency of order two and absolute anticommutativity. First property elaborates the 
fermionic nature of the (anti-)BRST symmetries whereas latter one insures that BRST and anti-BRST transformations 
are linearly independent of each other.  The anti-BRST symmetry is just not an artifact rather it plays an instrumental role in providing the geometrical 
interpretation of the superfield approach to BRST formalism \cite{bt1,bt2,del}. 
It is also useful in the investigation of perturbative renormalizability of Yang-Mills theory 
\cite{ore,alwa,hwa}. Thus, it is of utmost interest to study the (anti-)BRST invariant theory.

In our earlier work, we have shown that, in addition to the above fermionic (anti-)BRST symmetries, the nilpotent and absolutely 
anticommuting (anti-)co-BRST symmetries {\it also} exist for the Abelian $p$-form ($p =1, 2,3 $) gauge theories in a specific 
$D = 2p$-dimensions of spacetime within the framework of BRST formalism \cite{rpm1} where $D$ is the dimensionality of spacetime and 
$p$ denotes the degree of the differential form.  One of the key differences between the (anti-)BRST and (anti-)co-BRST symmetries is 
that the former symmetries leave the kinetic term invariant whereas under the latter symmetries the gauge--fixing term remains invariant. 
The appropriate anticommutators among the fermionic symmetries lead to a unique bosonic symmetry in the theory.
These fermionic and bosonic symmetries (and corresponding charges) 
provide the physical realizations of the de Rham cohomological operators of differential geometry  whereas discrete symmetry plays the 
role of Hodge duality $(*)$ operation (see, e.g. \cite{rpm1,rpm2,rpm3} for details). In fact, we have conjectured that in $D = 2p$-dimensions of spacetime, 
any arbitrary Abelian $p$-form gauge theory  ($p = 1, 2, 3,...$) provides a field-theoretic model for the Hodge theory within the framework of BRST formalism
\cite{rpm1}. Furthermore, we point out that the $(0 + 1)D$ rigid rotor and Christ--Lee model  in the context of  BRST formalism are shown to be the examples 
of  Hodge theory \cite{rpm3,ks}.

Christ--Lee (CL) model is one of the simplest examples of gauge-invariant system that is described by a singular Lagrangian. Physically, 
CL model represents a particle moving in plane with some specific constraints \cite{cl}. To be more specific, CL model is endowed with  
two first-class constraints in the language of Dirac's classification scheme of constraints \cite{dira,sund}. CL model has been 
well studied at the classical and quantum level in many different ways \cite{co1,co2,bl}. The gauge group of CL model is analogous 
to the quantum electrodynamics (QED) with a local gauge parameter varying as an arbitrary function of time. This simple physical 
system has been quantized by exploiting the usual canonical formalism with some specific gauge choices (e.g. temporal and/or Coulomb 
gauge conditions) \cite{cl}. This model has also been quantized by exploiting the BRST formalism \cite{kul}.

In our earlier 
work, we have shown that besides the usual off-shell nilpotent (anti-)BRST transformations there  also exist (anti-)co-BRST symmetries 
for CL model. Further, it has been explicitly shown that in addition to above fermionic transformations, a unique bosonic symmetry transformation is 
also present for this model within the framework of BRST formalism \cite{ks}. We have shown that these transformations 
(and corresponding conserved charges) obey an algebra which is exactly similar to the algebra satisfied by the de Rham cohomological 
operators ($d,\, \delta,\, \Delta$) \cite{egh,muk}. Thus, we have been able to show that CL model is a simple toy model for the 
Hodge theory \cite{ks}.

As far as the fermionic (anti-)BRST and (anti-)co-BRST symmetries are concerned, their geometrical origin become transparent and clear in the 
superfield formulation \cite{bt1,bt2,del}. Bonora-Tonin superfield approach to BRST formalism is a geometrically intuitive method 
where the key properties associated with the (anti-)BRST symmetry transformations find their geometrical origin in the language of 
Grassmannian translational generators in an elegant manner \cite{bt1,bt2}. In this formalism, a $D$-dimensional Minkowskian manifold 
is generalized to the ($D, 2$)-dimensional supermanifold. The latter is  parametrized by the superspace coordinates
$Z^M = (x^\mu, \eta, \bar\eta)$  where  $x^\mu$ $(\mu = 0, 1,..., D-1)$ are the bosonic coordinates  and ($\eta, \bar\eta$) 
are a pair of Grassmannian variables obeying nilpotency and anticommutativity properties 
(i.e. $\eta^2 = \bar\eta = 0,\; \eta\bar\eta + \bar\eta \eta = 0$). The superspace formalism, in general, allows superfields in a given field-theoretic model.  
One of the simplest examples is the Abelian 1-form gauge theory in 4D of spacetime. In the superfield approach to BRST formalism \cite{bt1,bt2,del}, 
we define a super 1-form
connection ${\cal A}^{(1)} = dZ^M {\cal A}_M(x,\eta,\bar\eta) \equiv dx^\mu {\cal A}_\mu + d\eta \,\bar {\cal C} + d \bar\eta \,{\cal C}$ 
and super exterior derivative $\tilde d = dZ^M \partial_M = dx^\mu \partial_\mu + d\eta\, \partial_\eta + d\bar\eta \,\partial_{\bar\eta}$ on the supermanifold corresponding to the $4D$ ordinary 1-form $A^{(1)} = dx^\mu\, A_\mu(x)$ and exterior derivative $d = dx^\mu\,\partial_\mu $. 
Here the superfields ${\cal A}_\mu(x, \eta, \bar\eta)$, ${\cal C}(x, \eta, \bar\eta)$ and $\bar {\cal C}(x, \eta, \bar\eta)$, as the supermultiplets of super 1-form, are the generalization of the 
gauge field $A_\mu(x)$, ghost field $C(x)$ and anti-ghost field $\bar C(x)$, respectively.     
Now we expand these superfields along the Grassmannian directions $\eta$ and $\bar\eta$ with the help of other secondary fields.
By exploiting the power and strength of {\it horizontality condition} (i.e. $\tilde d {\cal A}^{(1)} \equiv d A^{(1)}$), we precisely determine all 
the secondary fields in terms of the dynamical/auxiliary fields of the (anti-)BRST invariant theory. 
We point out that, the CL model is a $1D$ model where the dynamical variables
(as the generalized coordinates) are only the function of time-evolution parameter. We utilize here the sanctity of 
the above superfield approach for the present CL model.

For the interacting theories, a more powerful method known as  ``augmented" version of the superfield approach has 
been developed  where, in addition to the horizontality condition, the gauge-invariant restrictions are also implemented for 
the derivation of the complete set of proper (anti-)BRST transformations \cite{rpm4,rpm5,gk1,gk2}. 
In our present study, we shall utilize the power and strength of the superfield formalism to derive the off-shell nilpotent and absolutely anticommuting
(anti-)BRST as well as (anti-)co-BRST symmetry transformations for the $(0 + 1)$-dimensional CL model. In this approach, 
we have to go beyond the celebrated (dual-)horizontality condition to derive the proper (anti-)BRST and (anti-)co-BRST for all the 
dynamical variables present in the model. In fact, in addition to the (dual-)horizontality conditions, we use the gauge and (anti-)co-BRST invariant 
restrictions.

The contents of our present endeavour  are as follow. In section 2, we briefly  discuss about the CL model and 
associated local gauge symmetry. We also discuss about the supersymmetric type global (anti-)BRST and (anti-)co-BRST 
symmetry transformations (and corresponding conserved charges). Section 3 is devoted to the derivation of the off-shell 
nilpotent and absolutely anticommuting (anti-)BRST symmetry transformations with the help of ``augmented'' 
supervariable approach. Section 4 deals with the derivation of the proper (anti-) co-BRST transformations 
where the (anti-)co-BRST restrictions are used, in addition to the dual-horizontality condition. We capture 
the (anti-)BRST and (anti-)co-BRST invariances of the Lagrangian within the framework of supervariable 
approach in section 5. In our section 6, we show the nilpotency and anticommutativity properties of the 
(anti-)BRST and (anti-)co-BRST transformations (and corresponding generators) in terms of the translational 
generators along the directions of Grassmannian variables. Finally, in section 7, we provide the concluding remarks.


\section {Preliminaries: Christ--Lee model and associated symmetries}

 
We start off with the first-order as well as gauge-invariant Lagrangian of the $(0 + 1)$-dimensional Christ--Lee (CL) model as given by \cite{cl,co2,kul}
\begin{eqnarray}
&& L_f = \dot r \,p_r + \dot \theta\, p_\theta - \frac{1}{2}\, p^2_r - \frac{1}{2 r^2}\, p^2_\theta - z \,p_\theta - V(r),
\end{eqnarray} 
where $r, \,\theta$ are the generalized plane polar coordinates and $p_r, \, p_\theta$ are the corresponding canonical momenta, respectively. 
The variable $z$ is another generalized coordinate and $V(r)$ is the potential bounded from below. Under the following 
continuous gauge symmetry transformations 
\begin{eqnarray}
&& \delta\, z = \dot \chi(t), \qquad  \delta \,\theta = \chi(t), \qquad  \delta[r,\, p_r,\, p_\theta] =0,
\end{eqnarray}
where $\chi(t)$ is an infinitesimal local gauge parameter, the Lagrangian $L_f$ remains invariant (i.e., $\delta L_f = 0 $).

The (anti-)BRST invariant Lagrangian for the Christ--Lee model that incorporates the gauge--fixing term  and Faddeev--Popov (anti-)ghost variables
can be written as \cite{ks,kul}
\begin{eqnarray}
L = \dot r \,p_r + \dot \theta\, p_\theta - \frac{1}{2}\, p^2_r - \frac{1}{2 r^2}\, p^2_\theta - z \,p_\theta - V(r) 
+  \frac{1}{2}\,b^2 +  b (\dot z + \theta)  - i\,\dot {\bar C}\, \dot C + i \, \bar C \, C,
\end{eqnarray}
where $b$ is the Nakanishi--Lautrup type auxiliary variable and $(\bar C)C$ are the Faddeev--Popov (anti-)ghost variables
(with $C^2 = \bar C^2 = 0, C\bar C + \bar C C = 0$) having ghost numbers $(-1)+1$, respectively. 
The Lagrangian (3) respects the off-shell nilpotent ($s^2_{(a)b} = 0,\; s^2_{(a)d} = 0$) and absolutely 
anticommuting ($s_b\, s_{ab} + s_{ab}\, s_b = 0,\; s_d\, s_{ad} + s_{ad}\, s_d = 0$) (anti-)BRST  ($s_{(a)b}$)
and (anti-)co-BRST ($s_{(a)d}$) symmetry transformations. These continuous symmetries are listed as follows \cite{ks}
\begin{eqnarray}
&& s_b z = \dot C, \quad s_b \theta = C, \quad s_b \bar C = i\,b, \quad s_b[r, \,p_r,\, p_\theta,\, b, \,C] = 0, \nonumber\\
&& s_{ab} z = \dot {\bar C}, \quad s_{ab} \theta = \bar C, \quad s_{ab}\, C = - i b, \quad s_{ab}[r, \,p_r,\, p_\theta,b, \, \bar C] = 0, \nonumber\\
&&\nonumber\\
&& s_d\, z = \bar C, \quad s_d\, \theta = - \dot{\bar C}, \quad s_d \,C =  i\, p_\theta, \quad s_d\,[r,\, p_r,\, p_\theta,\, b,\, \bar C] = 0,\nonumber\\
&& s_{ad}\, z = C, \quad s_{ad}\, \theta = - \dot{C}, \quad s_{ad} \, \bar C = - i\, p_\theta, \quad s_{ad}\,[r,\, p_r,\, p_\theta,\, b,\, C] = 0.
\end{eqnarray}
One can readily check that under the above symmetry transformations $L$ remains quasi-invariant \cite{ecg}. To be more precise,
the Lagrangian transforms to a total time derivative under the above continuous and nilpotent symmetry transformations, namely; 
\begin{eqnarray}
&& s_b \,L = \frac{d}{dt}\big(b\, \dot C\big), \;\,\qquad\qquad s_{ab} \,L = \frac{d}{dt}\big(b\, \dot {\bar C}\big), \nonumber\\ 
&& s_d \,L = - \frac{d}{dt}\big(p_\theta\, \dot {\bar C}\big), \quad\qquad s_{ad} \,L = - \frac{d}{dt}\big(p_\theta\, \dot C\big). 
\end{eqnarray}
As a consequence, the action integral $S = \int dt L$ remains invariant under the (anti-)BRST and (anti-)co-BRST transformations. 
According to the Noether theorem, the invariance of the action under the above continuous symmetry transformations leads to the following 
conserved charges \cite{ks}: 
\begin{eqnarray}
&& Q_b = b\, \dot C + p_\theta \, C \,\equiv\, b \, \dot C - \dot b \, C, 
\qquad Q_{ab} = b\, \dot {\bar C} + p_\theta\, \bar C \, \equiv\, b \, \dot {\bar C} - \dot b \, {\bar C},  \nonumber\\
&& Q_d = b\, \bar  C - p_\theta \, \dot {\bar C} \,\equiv\, b \, \bar C + \dot b \, \dot {\bar C},
 \qquad Q_{ad} = b\,  C - p_\theta\, \dot C \,\equiv\,  b \, C + \dot b \, \dot C,
\end{eqnarray}
where on the r.h.s.,  we have used the equation of motion $p_\theta = - \dot b$ that has been derived from $L$. 
These conserved charges are the generators of the corresponding 
symmetry transformations. It is also to be noted that these charges are nilpotent of order two (i.e., $Q^2_{(a)b} = 0, \;Q^2_{(a)d} = 0$) and 
anticommuting ($Q_b\, Q_{ab} + Q_{ab}\, Q_b = 0, \; Q_d\, Q_{ad} + Q_{ad}\, Q_d = 0$) in nature.


\section{Off-shell nilpotent (anti-)BRST symmetries: supervariable approach}


We lay emphasis on the fact that the variable $z(t)$ behaves like a gauge variable \cite{nim} because, under the gauge transformations, 
it transform as $\delta z(t) = \dot\chi(t)$. For example, in QED,  the temporal component $A_0(x, t)$ of vector gauge field transforms as 
$\delta A_0 = \dot \Lambda(x,t)$ under the $U(1)$ gauge transformation where $\Lambda(x, t)$ is a local gauge parameter. Thus, we define the exterior
derivative $d$ (with $d^2 = 0$) and 1-form connection $Z^{(1)}$ on $(0+1)$-dimensional space parameterized only by (bosonic) time evolution parameter $t$ as
(see, e.g. \cite{egh,muk})
\begin{eqnarray}
&& d = dt \,\partial_t, \qquad Z^{(1)} = dt\, z(t).
\end{eqnarray}
We note that $d\,Z^{(1)} =0$ because of the property of wedge product $(dt \wedge dt) = 0$. In order to derive the proper (anti-)BRST transformations, 
we generalize the exterior derivative and 1-form to their corresponding super exterior derivative ($\tilde d$) and super 1-form 
($\tilde Z^{(1)}$), respectively on the
$(1, 2)$-dimensional superspace parametrized by (bosonic) $t$ and a pair of Grassmannian variables $(\eta, \bar \eta)$ 
(with $\eta^2 =  \bar \eta^2 = 0, \; \eta\, \bar \eta + \bar \eta\, \eta = 0$) as given by (see, e.g. \cite{bt1,bt2} for details)
\begin{eqnarray}
d \to \tilde d &=& dt\, \partial_t + d \eta\, \partial_{\eta} + d \bar \eta\, \partial_{\bar \eta}, \qquad  (\tilde d^2 = 0),\nonumber\\
Z^{(1)} \to \tilde Z^{(1)} &=& dt\, {\cal  Z}(t, \eta, \bar \eta) + d \eta \, \bar {\cal F}(t, \eta, \bar \eta) + d \bar \eta \, {\cal F}(t, \eta,  \bar \eta), 
\end{eqnarray} 
where $\partial_{\eta} = \partial/\partial \eta$ and $\partial_{\bar \eta} = \partial/\partial \bar\eta$ are the Grassmannian derivatives 
(with $\partial^2_{\eta} =  \partial^2_{\bar \eta} = 0, \; \partial_{\eta}\,\partial_{\bar \eta}+ \partial_{\bar \eta}\,\partial_{\eta} = 0$) corresponding to 
the variables $\eta$ and $\bar \eta$, respectively. The super multiplets as the 
components of super 1-form can be expanded along the directions of Grassmannian variables ($\eta, \bar \eta$) as follows 
\begin{eqnarray}
{\cal Z}(t, \eta, \bar \eta) &=& z(t) + \eta\, \bar f_1(t) + \bar \eta \, f_1(t) + i \eta\, \bar \eta\, B(t), \nonumber\\
{\cal F}(t, \eta, \bar \eta) &=& C(t) + i\,\eta\, \bar b_1(t) + i\,\bar \eta \, b_1(t) + i \eta\, \bar \eta\, s(t), \nonumber\\
\bar {\cal F} (t, \eta, \bar \eta) &=& \bar C(t) + i\,\eta\, \bar b_2(t) + i\,\bar \eta \, b_2(t) + i \eta\, \bar \eta\, \bar s(t), 
\end{eqnarray}
where $B, \, b_1, \,\bar b_1,\, b_2, \bar b_2$ are the bosonic secondary  variables and $f_1,\, \bar f_1,\, s,\, \bar s$ are 
the fermionic secondary variables. We shall determine these secondary variables in terms of the basics and auxiliary variables by 
exploiting the following horizontality condition 
\begin{eqnarray}
\tilde d\, \tilde Z^{(1)} = d \, Z^{(1)}.
\end{eqnarray}
The  horizontality condition is also known as {\it ``soul-flatness''} condition where $t$ is a  body coordinate and 
($\eta, \, \bar \eta$) are the soul coordinates \cite{noji}. The horizontality or soul-flatness condition  implies that the r.h.s. would remain 
independent of the soul coordinates when it is generalized onto $(1, 2)$D supermanifold. The l.h.s. of equation (10), 
in full blaze of glory, can be written as 
\begin{eqnarray}
\tilde d \tilde Z^{(1)} &=& (dt \wedge d \eta)\left(\partial_t \bar {\cal F} - \partial_\eta {\cal  Z}\right) 
+ (dt \wedge d \bar \eta)\left(\partial_t {\cal F} - \partial_{\bar \eta} {\cal Z}\right) \nonumber\\
&-& (d\eta \wedge d \bar \eta)\left(\partial_\eta {\cal F} + \partial_{\bar \eta} \bar {\cal F} \right)
- (d\eta \wedge d \eta)\left(\partial_\eta \bar {\cal F} \right)
- (d \bar \eta \wedge d \bar \eta)\left(\partial_{\bar \eta} {\cal F} \right). \qquad
\end{eqnarray}
Exploiting (10) and (11) with  the help of (9), we obtain the following interesting relationships 
among the basic and secondary variables; namely,
\begin{eqnarray}
&& f_1 = \dot C, \qquad \bar f_1 = \dot {\bar C}, \qquad b_2 + \bar b_1 = 0, \qquad B  = \dot b,\nonumber\\
&& b_1 = 0, \qquad \bar b_2 = 0, \qquad s =0, \qquad \bar s = 0,
\end{eqnarray}
where we have made the choice $b_2 = - \bar b_1 = b$ for the Nakanishi--Lautrup type auxiliary variable. Substituting the values 
of secondary variables from (12) in (9), we yield the following expressions for the supervariables: 
\begin{eqnarray}
{\cal Z}^{(h)}(t, \eta, \bar \eta) &=& z(t) + \eta\, \dot {\bar C}(t) + \bar \eta \, \dot C(t) + i \eta\, \bar \eta\, \dot b(t), \nonumber\\
{\cal F}^{(h)}(t, \eta, \bar \eta) &=& C(t) - i\,\eta\, b(t), \nonumber\\
\bar {\cal F}^{(h)} (t, \eta, \bar \eta) &=& \bar C(t) + i\,\bar \eta \, b(t), 
\end{eqnarray}  
where the superscript $(h)$ on supervariables implies that the super expansions of supervariables obtained  after the application of 
horizontality condition (10).

At this juncture, we lay emphasis on the fact that the quantity $(z - \dot \theta)$ remains invariant under the gauge transformations (2). 
Thus, it would also be independent of the Grassmannian variables when we generalize it onto $(1, 2)$-dimensional superspace. 
This gauge-invariant quantity will serve our purpose to derive the off-shell nilpotent (anti-)BRST transformations for $\theta$ variable 
\cite{rpm4,rpm5,gk1}. 
In the language of differential form, we can write this gauge-invariant quantity as follows   
\begin{eqnarray}
Z^{(1)} - d \, \theta^{(0)} = dt \big(z(t) - \partial_t \theta(t) \big),
\end{eqnarray}  
which is clearly a 1-form object. Here $\theta^{(0)} = \theta$ is a zero-form.  Now, we generalize this 1-form object onto $(1, 2)$-dimensional 
supermanifold as 
\begin{eqnarray}
\tilde Z^{(1)} - \tilde d \, \tilde \theta^{(0)} = Z^{(1)} - d \, \theta^{(0)},
\end{eqnarray}
where the super zero-form $\tilde \theta^{(0)}$ is defined in the following fashion: 
\begin{eqnarray}
\tilde \theta^{(0)} &=& \Theta(t, \eta, \bar \eta)\nonumber\\
&=& \theta(t) + \eta\, \bar f_2 + \bar \eta f_2 + i\, \eta \bar \eta \bar B.
\end{eqnarray}
In the above, ${\bar B}$  is a bosonic secondary variable whereas $f_2, \bar f_2$ are the fermionic secondary variables.  
The l.h.s. of (15) can be  explicitly written as 
\begin{eqnarray}
\tilde Z^{(1)} - \tilde d \, \tilde \theta^{(0)} = dt \big[{\cal Z}^{(h)} - \partial_t \Theta \big] + d \eta \big[\bar {\cal F}^{(h)} 
- \partial_\eta \Theta \big] + d \bar \eta \big[{\cal F}^{(h)} - \partial_{\bar \eta} \Theta \big].
\end{eqnarray}
Using (15) and (17) together with (13), we obtain the precise value of the secondary variables
\begin{eqnarray}
&& f_2 = C, \qquad \bar f_2 = \bar C, \qquad \bar B = b.
\end{eqnarray}
Furthermore, we point out that the dynamical variables $r,\, p_r$ and $p_\theta$ are also gauge-invariant as one can see from (2). 
These gauge-invariant variables would also remain unaffected by the presence of Grassmannian variables. As a result, 
we obtain the following super expansions; namely, 
\begin{eqnarray}
\Theta^{(h)}(t, \eta, \bar \eta) &=& \theta(t) + \eta\, \bar C + \bar \eta \, C + i \eta \bar \eta\, b, \nonumber\\ 
{\cal R}^{(h)}(t, \eta, \bar \eta) &=& r(t),\nonumber\\
{\cal P}^{(h)}_r (t, \eta, \bar \eta) &=& p_r (t), \nonumber\\
{\cal P}^{(h)}_{\theta} (t, \eta, \bar \eta) &=& p_\theta (t).
\end{eqnarray}
It is to be noted that if we look carefully at the super-expansions  given in equations (13) and (19), we can easily find out the proper (anti-)BRST transformations for all the dynamic variables. 
In fact, the BRST and anti-BRST transformations can be obtained for any generic dynamical variable $\phi(t)$  from its corresponding 
supervariable $\Phi^{(h)}(t, \eta, \bar \eta)$ in the following manner:   
\begin{eqnarray}
&& s_b\, \phi(t) = \frac{\partial}{\partial \bar \eta}\, \Phi^{(h)}(t, \eta, \bar \eta)\Big|_{\eta = 0}, \qquad
s_{ab}\, \phi(t) = \frac{\partial}{\partial \eta}\, \Phi^{(h)}(t, \eta, \bar \eta)\Big|_{\bar \eta = 0},\nonumber\\
&& s_b \,s_{ab} \,\phi(t) =  \frac{\partial}{\partial \bar \eta} \,\frac{\partial}{\partial \eta}\, \Phi^{(h)}(t, \eta, \bar \eta).
\end{eqnarray}
Using the above equations, we obtain the off-shell nilpotent and absolutely anticommuting (anti-)BRST symmetry transformations as listed in (4) 
\cite{ks}. However, the (anti-)BRST transformations for the Nakanishi--Lautrup variable $b$ have been derived from the requirements 
of nilpotency and anticommutativity of the (anti-)BRST transformations.

Exploiting the basic tenets  of BRST formalism, we can write the Lagrangian (3) in three different ways by using the (anti-)BRST $(s_{(a)b})$ 
transformations as 
\begin{eqnarray}
L &=&  \dot r \,p_r + \dot \theta\, p_\theta - \frac{1}{2}\, p^2_r - \frac{1}{2 r^2}\, p^2_\theta - z \,p_\theta - V(r) 
- s_b \bigg[i\, \bar C\left(\dot z + \theta + \frac{b}{2}\right)\bigg]\nonumber\\
&\equiv&  \dot r \,p_r + \dot \theta\, p_\theta - \frac{1}{2}\, p^2_r - \frac{1}{2 r^2}\, p^2_\theta - z \,p_\theta - V(r)
+ s_{ab} \bigg[i \, C\left(\dot z + \theta + \frac{b}{2}\right)\bigg] \nonumber\\
&\equiv&  \dot r \,p_r + \dot \theta\, p_\theta - \frac{1}{2}\, p^2_r - \frac{1}{2 r^2}\, p^2_\theta - z \,p_\theta - V(r)  
+ s_b\, s_{ab} \bigg[\frac{i}{2} \,\big(z^2 - \theta^2 \big) - \frac{1}{2}\,\bar C\, C\bigg],
\end{eqnarray}
modulo a total time derivative term. It is clear from the above that, due to the nilpotency property $(s^2_{(a)b} = 0)$ of  $s_{(a)b}$,
the (anti-)BRST invariance of $L$ can now be proven in a simple and straightforward manner.


\section{Off-shell nilpotent (anti-)co-BRST symmetries: supervariable approach}


In this section,  we shall derive the off-shell nilpotent  (i.e., $s^2_{(a)d} = 0$) and absolutely anticommuting
(i.e., $s_d\,s_{ad} + s_{ad} \, s_d = 0$)  (anti-)co-BRST symmetry transformations ($s_{(a)d}$). We accomplish 
this goal by exploiting the power and strength of the dual-horizontality condition together with (anti-)co-BRST 
invariant restrictions. The action of co-exterior derivative $\delta = *\, d\,*$ (with $\delta^2 = 0$) on 1-form 
$Z^{(1)}$ yields 
\begin{eqnarray}
&& \delta \,Z^{(1)}  = * \,d \,* Z^{(1)} = \dot z(t),
\end{eqnarray}
where ($*$) is the Hodge duality operation defined on $(0+1)$-dimensional manifold. The gauge--fixing term $(\dot z + \theta)$, 
which remains invariant under (anti-)co-BRST symmetries, can  be written in the following fashion:   
\begin{eqnarray}
&& \delta \,Z^{(1)} + \theta^{(0)} = \dot z(t) + \theta (t).
\end{eqnarray}
The invariance of gauge--fixing term under the (anti-)co-BRST transformations can be captured in the following (anti-)co-BRST
invariant restriction \cite{rpm2,rpm5}
\begin{eqnarray}
&& \star\, \tilde d\, \star \,\tilde Z^{(1)}  + \Theta   = *\, d\, * \,Z^{(1)} + \theta,
\end{eqnarray}
which tells us that the r.h.s. is independent of the Grassmannian variables $\eta$ and $\bar \eta$ when we generalize it onto
$(1, 2)$D supermanifold.
In the above, the super co-exterior derivative $\tilde \delta = \star \,\tilde d \,\star$ (with ${\tilde \delta}^2 = 0$) 
and the Hodge duality $(\star)$ operation are defined onto $(1, 2)$-dimensional supermanifold.  
In terms of the supervariables,  one can simplify (24) as   
\begin{eqnarray}
\left(\partial_t {\cal Z} + \partial_\eta \bar {\cal F} + \partial_{\bar \eta} {\cal F}\right) 
+ S^{\eta\eta} \, \partial_\eta {\cal F} + S^{\bar \eta \bar \eta}\, \partial_{\bar \eta} \bar {\cal F} + \Theta 
= \dot z + \theta,
\end{eqnarray}
where $S^{\eta \eta}$ and $S^{\bar \eta \bar \eta}$ are symmetric in $\eta$ and $\bar \eta$. 
In the above equation, we have used the following mathematical definitions defined on $(1, 2)$-dimensional supermanifold  \cite{rpm2,rpm5}
\begin{eqnarray}
&& \star\, dt = (d\eta \wedge d \bar \eta),  \qquad \qquad  \star\, (dt \wedge d\eta \wedge d \bar \eta) = 1,  \nonumber\\
&& \star\, d\eta = (dt \wedge d \bar \eta),  \qquad \qquad \star\, (dt \wedge d\eta \wedge d \eta) = S^{\eta \eta},  \nonumber\\
&& \star \,d\bar \eta  = (dt \wedge d \eta), \qquad \qquad  \star \,(dt \wedge d\bar \eta \wedge d \bar \eta) = S^{\bar \eta \bar \eta}, \nonumber\\
&&(dt \wedge dt \wedge d \eta ) = 0,         \qquad \qquad  (d\eta \wedge d \eta \wedge d \bar \eta ) = 0,\nonumber\\
&& (d\eta \wedge d \bar \eta \wedge d \bar \eta ) = 0.
\end{eqnarray} 
From equation (25), we finally yield the following relationships: 
\begin{eqnarray}
&& \bar f_2 = - \dot {\bar f}_1, \qquad f_2 = - \dot f_1, \qquad \bar B = - \dot B,  \qquad s = 0,\nonumber\\
&& b_1 = - \bar b_2, \qquad \bar b_1 = 0, \qquad b_2 = 0, \qquad \bar s = 0.
\end{eqnarray}
Substituting these relationships in (9), we obtain the following super-expansions for the supervariables
\begin{eqnarray}
\Theta^{(r)} (t, \eta, \bar \eta) &=& \theta(t) - \eta\, \dot {\bar f}_1 (t)  - \bar \eta\, \dot f_1 (t)  - i \eta \bar \eta\, \dot B (t) , \nonumber\\
{\cal F}^{(r)}(t, \eta, \bar \eta) &=& C(t)  + i \bar \eta\, {\cal B}(t) , \nonumber\\
\bar {\cal F}^{(r)}(t, \eta, \bar \eta) &=& \bar C(t)  - i \eta\, {\cal B}(t),
\end{eqnarray}
where we have chosen $b_1 = - \bar b_2 = {\cal B}$ for our algebraic convenience and the superscript $(r)$ 
denotes the reduced form of the supervariables [cf. (9) and (16)].

It is clear that we have not obtained the super-expansions in terms of the basic variables of the present theory. 
In fact, the coefficients of $\eta, \bar \eta$, and $\eta \bar \eta$ in the expression of supervariables are still unknown. 
Thus, to accomplish this goal,  we invoke the (anti-)co-BRST invariant restrictions on the dynamical variables. These restrictions are
(see, e.g. \cite{rpm5,gk2} for details) 
\begin{eqnarray}
s_{(a)d} \big[z\, p_\theta - i\, \bar C\, C\big] = 0, \qquad\qquad s_{(a)d} \big[\theta \,p_\theta - i\, \dot {\bar C}\, C\big] = 0.
\end{eqnarray}
We demand that these (anti-)co-BRST invariant restrictions would remain intact when we generalize them onto $(1, 2)$-dimensional supermanifold. 
As a result, we can write 
\begin{eqnarray}
&&{\cal Z} {\cal P}_\theta - i \bar {\cal F}^{(r)} {\cal F}^{(r)} = z p_\theta - i \bar C C, \nonumber\\
&& \Theta^{(r)} {\cal P}_\theta - i\, \partial_t {\bar {\cal F}}\,^{(r)}  {\cal F}^{(r)} = \theta p_\theta - i \dot {\bar C} C.
\end{eqnarray}
Exploiting the equations (28) and (30), we yield the following relationships:   
\begin{eqnarray}
&& p_\theta \bar f_1 - {\cal B} C = 0, \qquad p_\theta f _1 - {\cal B} \bar C = 0, \qquad B p_\theta - {\cal B}{\cal B} = 0, \nonumber\\ 
&& p_\theta \dot {\bar f}_1 - \dot {\cal B} C = 0, \qquad p_\theta \dot f _1 - {\cal B} \dot {\bar C} = 0, 
\qquad \dot B p_\theta - \dot {\cal B}{\cal B} = 0.
\end{eqnarray}
Here we again emphasis on the fact that the restrictions in (29) are not enough to determine the precise values of secondary variables. We further note that 
$s_d (z \bar C) = 0, \; s_{ad} (z C) = 0$. These co-BRST and anti-co-BRST invariant restrictions would remain independent of $\eta$ and $\bar \eta$. 
The generalization of these restrictions onto $(1, 2)$-dimensional manifold yield the following interesting relationships:   
\begin{eqnarray}
{\cal Z} \bar {\cal F}^{(r)} = z \bar C \Rightarrow 
\begin{cases}
f_1\, \bar C = 0,\\
B\, \bar C - f_1 {\cal B} = 0,\\
\bar f_1\, \bar C - i z {\cal B} = 0,
\end{cases}
\qquad \text{and} \quad
{\cal Z}  {\cal F}^{(r)} = z C \Rightarrow
\begin{cases}
\bar f_1\, C = 0,\\
B\, C -  \bar f_1 {\cal B} = 0,\\
\bar f_1\,  C + i z {\cal B} = 0.
\end{cases} 
\end{eqnarray} 
It is clear that the relations $f_1\,\bar C = 0$ and $\bar f_1\, C = 0$ fix the value of secondary variables  $f_1$ and $\bar f_1$ as 
$f_1 \propto \bar C$ and $\bar f_1 \propto C$ \cite{rpm5,gk2}. The simplest solutions that satisfy the relationships appear 
in equations (31) and (32) are 
\begin{eqnarray}
f_1 = \bar C, \qquad \bar f_1 = C, \qquad {\cal B} = p_\theta = B. 
\end{eqnarray} 
As a consequence, we obtain the precise values of the secondary variables in terms of the basic and auxiliary variables. 
Further, it is to be noted that the dynamical variables $r$, $p_r$, and $p_\theta$ are (anti-)co-BRST invariant and, 
thus, the supervariables corresponding to them would remain independent of the Grassmannian variables.  The supervariables now have 
the following expansions along the Grassmannian directions as follows:
\begin{eqnarray}
{\cal Z}^{(d)}(t, \eta, \bar \eta) &=& z(t) + \eta\, C(t) + \bar \eta \, \bar C(t) + i \eta\, \bar \eta\, p_\theta(t), \nonumber\\
\Theta^{(d)}(t, \eta, \bar \eta) &=& \theta(t) - \eta\, \dot C(t) - \bar \eta \, \dot {\bar C}(t) - i \eta\, \bar \eta\, \dot p_\theta(t), \nonumber\\
{\cal F}^{(d)}(t, \eta, \bar \eta) &=& C(t) + i\,\bar \eta\, p_\theta(t), \nonumber\\
\bar {\cal F}^{(d)} (t, \eta, \bar \eta) &=& \bar C(t) - i\, \eta \, p_\theta(t), \nonumber\\
{\cal R}^{(d)}(t, \eta, \bar \eta) &=& r(t),\nonumber\\
{\cal P}^{(d)}_r (t, \eta, \bar \eta) &=& p_r (t), \nonumber\\
{\cal P}^{(d)}_{\theta} (t, \eta, \bar \eta) &=& p_\theta (t),
\end{eqnarray} 
where the superscript $(d)$ denotes that the above expressions for the supervariables obtained after the application of 
dual-horizontality conditions together with the (anti-)co-BRST invariant restrictions.      
Now, from the above expansions of the supervariables, we obtain the complete set of off-shell nilpotent and absolutely anticommuting 
(anti-)co-BRST symmetry transformations [cf. (4)] (see, Ref. \cite{gk2} for details). 
To be more specific, the co-BRST ($s_d$) and anti-co-BRST ($s_{ad}$) transformations 
for any generic variable can be obtained from their corresponding supervariable as
\begin{eqnarray}
&& s_d \phi(t) = \frac{\partial}{\partial \bar \eta}\, \Phi^{(d)}(t, \eta, \bar \eta)\Big|_{\eta = 0}, \qquad
s_{ad} \phi(t) = \frac{\partial}{\partial \eta}\, \Phi^{(d)}(t, \eta, \bar \eta)\Big|_{\bar \eta = 0},\nonumber\\
&& s_d \,s_{ad} \phi(t) =  \frac{\partial}{\partial \bar \eta} \,\frac{\partial}{\partial \eta}\, \Phi^{(d)}(t, \eta, \bar \eta).
\end{eqnarray}
In other words, the co-BRST symmetry transformation $(s_d)$ is equivalent to the translation of the generic supervariable 
$\Phi^{(d)}(t, \eta, \bar \eta)$ along $\bar \eta$-direction while keeping $\eta$-direction fixed. Similarly,  
the anti-co-BRST transformation $(s_{ad})$ can be obtained by taking the translation of the generic supervariable 
$\Phi^{(d)}(t, \eta, \bar \eta)$ along $\eta$-direction while $\bar \eta$-direction 
remains intact.

Before we wrap this section, we point out that the total gauge--fixing terms $\frac{1}{2}\, b^2 + b\big(\dot z - \theta\big)$ remain invariant
under (anti-)co-BRST transformations.  Furthermore,
the three terms $\dot r\, p_r  - \frac{1}{2 r^2}\, p^2_\theta + V(r)$ do not transform under (anti-)co BRST transformations because 
the dynamical variables $r, \, p_r, p_\theta$ remain invariant under the off-shell nilpotent (anti-)co-BRST symmetry transformations (4). 
The rest of terms in $L$, we can write as the co-BRST exact term 
and anti-co-BRST exact term. As a consequence, the Lagrangian can be written in two different ways 
in terms of $s_d$ and  $s_{ad}$ as follows:
\begin{eqnarray}  
L &=& \dot r\, p_r  - \frac{1}{2 r^2}\, p^2_\theta + V(r) + \frac{1}{2}\, b^2 + b\big(\dot z - \theta\big) 
+ s_d \big[+\, i\, C \big(\dot z - \theta\big)\big] \nonumber\\
&\equiv&\dot r\, p_r  - \frac{1}{2 r^2}\, p^2_\theta + V(r) + \frac{1}{2}\, b^2 + b\big(\dot z - \theta\big) 
+ s_{ad} \big[-\, i\, \bar C \big(\dot z - \theta\big)\big],
\end{eqnarray}
modulo a total time derivative. It is now clear from the above that the (anti-)co-BRST invariance of $L$ 
can be proven in a simpler way because of the nilpotency property of the (anti-)co-BRST transformations.

  
\section{Invariance of Lagrangian}  

In this section, we capture the (anti-)BRST and (anti-)co-BRST invariances of the Lagrangian in terms of the 
Grassmannian translational generators $(\partial_\eta,\, \partial_{\bar \eta})$. To accomplish this goal, we generalize the total 
Lagrangian $(L)$ from ($0 +1)$-dimensional manifold to super Lagrangian (${\cal L}$) defined onto $(1, 2)-$dimensional supermanifold.

We note that the gauge-invariant (first-order) Lagrangian (1) can be generalized to super 
Lagrangian in terms of the supervariables (13) and (19) as 
\begin{eqnarray}
L_f \to {\cal L}_f &=& \dot r\, p_r + \dot \Theta^{(h)} \, p_\theta - \frac{1}{2}\, p^2_r - \frac{1}{2\,r^2}\,p^2_{\theta} 
- {\cal Z}^{(h)}\, p_\theta - V(r).
\end{eqnarray}
One can check that the super Lagrangian ${\cal L}_f$, defined onto (1, 2)-dimensional supermanifold,
is independent of the Grassmannian variables ($i.e., {\cal L}_f = L_f$) and 
this is the reason behind the invariance of $L_f$  under the (anti-)BRST transformations. This statement, mathematically, can be 
corroborated in terms of the translational generators along the Grassmannian directions as follows:
\begin{eqnarray}
\frac{\partial}{\partial \bar \eta} \, {\cal L}_f  = 0 & \Longleftrightarrow  & s_b L_f = 0, \nonumber\\
\frac{\partial}{\partial \eta} \, {\cal L}_f  = 0 &\Longleftrightarrow & s_{ab} L_f = 0.
\end{eqnarray} 
Similarly, the total Lagrangian $L$ onto $(1, 2)$-dimensional supermanifold can be written as 
\begin{eqnarray}
{\cal L} = {\cal L}_f +  \frac{1}{2}\,b^2 + b \big(\dot {\cal Z}^{(h)} + \Theta^{(h)} \big)  
- i \,{\dot {\bar {\cal F}}}^{(h)}\, \dot {\cal F}^{(h)} + i\,\bar {\cal F}^{(h)}\, {\cal F}^{(h)}.
\end{eqnarray}
The quasi-(anti-)BRST invariance of the total Lagrangian $L$ [cf. (5)] can be translated in terms of the above super Lagrangian and 
the Grassmannian derivatives as
\begin{eqnarray}
&& \frac{\partial}{\partial \bar \eta} \, {\cal L}\Big|_{\eta = 0} =  \frac{d}{dt}\big(b\, \dot C\big) 
\Longleftrightarrow s_b L = \frac{d}{dt}\big(b\, \dot C\big), \nonumber\\
&& \frac{\partial}{\partial \eta} \, {\cal L}\Big|_{\bar \eta = 0} = - \frac{d}{dt}\big(b\, \dot {\bar C}\big) 
\Longleftrightarrow s_{ab} L = - \frac{d}{dt}\big(b\, \dot {\bar C}\big).
\end{eqnarray}
Mention should be made here that the super Lagrangian (39), after a bit algebraic computation, leads to the Lagrangian 
(3) plus total time derivative terms which contain Grassmannian variables $(\eta, \bar \eta)$.
This is why, the actions of Grassmannian derivatives on (39) lead to total derivatives.

Furthermore, there are two more ways to write the super Lagrangian (39) in the language of supervariables. 
These are listed as follows 
\begin{eqnarray}
L \to {\cal L}  &=& {\cal L}_f+  \frac{\partial}{\partial \bar \eta}\bigg[- i \bar {\cal F}^{(h)}\left(\dot {\cal Z}^{(h)} + \Theta^{(h)} 
+ \frac{b}{2}\right)\bigg] \bigg|_{\eta = 0}\nonumber\\
&=& {\cal L}_f+ \frac{\partial}{\partial \eta}\bigg[i {\cal F}^{(h)}\left(\dot {\cal Z}^{(h)} + \Theta^{(h)} 
+ \frac{b}{2}\right)\bigg] \bigg|_{\bar \eta = 0}\nonumber\\
&=&  {\cal L}_f+ \frac{\partial}{\partial \bar \eta}\, \frac{\partial}{\partial \eta}\, \bigg[\frac{i}{2} \Big({\cal Z}^{(h)} \,{\cal Z}^{(h)} 
- \Theta^{(h)} \, \Theta^{(h)} \Big) - \frac{1}{2} \, \bar {\cal F}^{(h)}\, {\cal F}^{(h)}\bigg]. 
\end{eqnarray}
We notice that (anti-)BRST invariance of total Lagrangian $L$ can also be captured in terms of the Grassmannian derivatives as follows   
\begin{eqnarray}
\frac{\partial}{\partial \bar \eta} \, {\cal L} \Big|_{\eta = 0} = 0 \Leftrightarrow s_b L = 0, \qquad \quad
\frac{\partial}{\partial \eta} \, {\cal L} \Big|_{\bar \eta = 0} = 0 \Leftrightarrow s_{ab} L = 0,
\end{eqnarray}
where we have used the nilpotency property ($\partial^2_\eta  = \partial^2_{\bar \eta} = 0$) of the Grassmannian 
derivatives $\partial_\eta$ and $\partial_{\bar \eta}$. At this juncture, we point out that there is a little difference between (40) 
and (42). This happens because of the fact that we have discarded the total time derivative term while deriving the Lagrangian (3) from 
(21) (without any loss of generality). Because we know that the total time derivative term in the Lagrangian (or action)
 does not affect the dynamics of the system.

In an exactly similar fashion, one can also prove the (anti-)co-BRST invariance of the Lagrangian. For this purpose, we
generalize the Lagrangian (3) from an ordinary $1$-dimensional space  to the $(1, 2)$-dimensional superspace in terms of the supervariables (34) as 
\begin{eqnarray}
L \to {\cal L} &=& \dot r\,p_r + \dot \Theta^{(d)}\, p_\theta - \frac{1}{2}\,p^2_r - \frac{1}{2\,r^2}\,p^2_\theta
 - {\cal Z}^{(d)}\, p_\theta  \nonumber\\
&-&  V(r) +  \frac{1}{2}\,b^2 + b \big(\dot {\cal Z}^{(d)} + \Theta^{(d)} \big)  
- i \,{\dot {\bar {\cal F}}}^{(d)}\, \dot {\cal F}^{(d)} + i\,\bar {\cal F}^{(d)}\, {\cal F}^{(d)}.
\end{eqnarray}
Upon simplifying the above super Lagrangian, we note that it is independent of the Grassmannian variables 
($\eta$ and $\bar \eta$). In fact, it leads to the Lagrangian (3) modulo a total time derivative term. As a consequence, we yield
\begin{eqnarray}
&& \frac{\partial}{\partial \bar \eta} \, {\cal L}\Big|_{\eta = 0} = - \frac{d}{dt}\big(p_\theta\, \dot {\bar C}\big) 
\Longleftrightarrow s_d L = - \frac{d}{dt}\big(p_\theta\, \dot {\bar C}\big), \nonumber\\
&& \frac{\partial}{\partial \eta} \, {\cal L}\Big|_{\bar \eta = 0} = - \frac{d}{dt}\big(p_\theta\, \dot C\big) 
\Longleftrightarrow s_{ad} L = - \frac{d}{dt}\big(p_\theta\, \dot C\big).
\end{eqnarray}
which are consistent with the equations given in (5).

As we already know that the total gauge--fixing term $\frac{b^2}{2} + b(\dot z + \theta)$ are (anti-)co-BRST invariant [cf. (4)]. 
Thus, the total gauge--fixing super Lagrangian 
\begin{eqnarray}
&& {\cal L}_{GF}  = \frac{1}{2}\,b^2 + b \, \big(\dot {\cal Z}^{(d)} + \Theta^{(d)}\big),  
\end{eqnarray}
as one can easily check, is independent of the Grassmannian variables. In fact, we have 
\begin{eqnarray}
&& \frac{\partial}{\partial \eta}\,{\cal L}_{GF}  = 0, \qquad  \frac{\partial}{\partial \bar \eta}\,{\cal L}_{GF} = 0,
\end{eqnarray}  
which reflect the fact that the total gauge--fixing terms remain invariant under the off-shell nilpotent (anti-)co-BRST transformations. 
Furthermore, the dynamical variables $r, p_r$ and $p_\theta$ remain invariant under the (anti-)co-BRST transformations. 
Thus, we can write the total super Lagrangian in two more different ways in terms of the supervariables (34) as   
\begin{eqnarray}
{\cal L}  = \dot r\,p_r - \frac{1}{2 r^2}\, p^2_\theta + V(r) + \frac{1}{2}\,b^2 +  b \, \big(\dot {\cal Z}^{(d)} 
+ \Theta^{(d)}\big) + \frac{\partial}{\partial \eta}\bigg[+ \,i\, {\cal F}^{(d)}\bigg({\cal Z}^{(d)} 
- \dot \Theta^{(d)}\bigg)\bigg] \bigg|_{\bar \eta = 0}\nonumber\\
\equiv \dot r\,p_r - \frac{1}{2 r^2}\, p^2_\theta + V(r) + \frac{1}{2}\,b^2 +  b \, \big(\dot {\cal Z}^{(d)} 
+ \Theta^{(d)}\big) + \frac{\partial}{\partial \bar \eta}\bigg[- i\, \bar {\cal F}^{(d)}\bigg({\cal Z}^{(d)} 
+ \dot \Theta^{(d)} \bigg)\bigg] \bigg|_{\eta = 0}.
\end{eqnarray}
It is clear from the above super Lagrangian that the (anti-)co-BRST invariance of the Lagrangian can now be 
proven in a simpler way due to the nilpotency property $(\partial^2_\eta = \partial^2_{\bar \eta} = 0)$ 
of the Grassmannian derivatives $(\partial_\eta, \, \partial_{\bar \eta})$.


\section{Nilpotency and absolute anticommutativity check} 

The (anti-)BRST as well as  (anti-)co-BRST symmetry transformations obey two key  properties: $(i)$ nilpotency of order two, and 
$(ii)$ absolute anticommutativity. The nilpotency property for any generic 
variable  can be translated into superspace in terms of the corresponding supervariable and Grassmannian translational generators as follows:  
\begin{eqnarray}
s^2_b \phi(t) = 0 &\Leftrightarrow & \frac{\partial}{\partial \bar\eta}\, \frac{\partial}{\partial \bar\eta}\, \Phi^{(h)}(t, \eta, \bar \eta) = 0, \nonumber\\
s^2_{ab} \phi(t) = 0 &\Leftrightarrow &  \frac{\partial}{\partial \eta}\, \frac{\partial}{\partial \eta}\, \Phi^{(h)}(t, \eta, \bar \eta) = 0, \nonumber\\
s^2_d \phi(t) = 0 &\Leftrightarrow & \frac{\partial}{\partial \bar\eta}\, \frac{\partial}{\partial \bar\eta}\, \Phi^{(d)}(t, \eta, \bar \eta) = 0, \nonumber\\
s^2_{ad} \phi(t) = 0 &\Leftrightarrow & \frac{\partial}{\partial \eta}\, \frac{\partial}{\partial \eta}\, \Phi^{(d)}(t, \eta, \bar \eta) = 0. \qquad
\end{eqnarray}
Similarly, the absolute anticommutativity property of the above nilpotent symmetry transformations can also be captured in 
terms of the supervariables and Grassmannian derivatives as given below:
\begin{eqnarray}
&& (s_b\, s_{ab} + s_{ab}\, s_b) \phi(t) = 0 \; \Leftrightarrow \; \bigg(\frac{\partial}{\partial \bar \eta}\, \frac{\partial}{\partial \eta} 
+ \frac{\partial}{\partial \eta}\, \frac{\partial}{\partial \bar \eta}\bigg) \, \Phi^{(h)}(t, \eta, \bar \eta) = 0, \nonumber\\
&& (s_d\, s_{ad} + s_{ad}\, s_d) \phi(t) =0 \; \Leftrightarrow \;  \bigg(\frac{\partial}{\partial \bar \eta}\, \frac{\partial}{\partial \eta} 
+ \frac{\partial}{\partial \eta}\, \frac{\partial}{\partial \bar \eta}\bigg) \, \Phi^{(d)}(t, \eta, \bar \eta) = 0,  
\end{eqnarray}
where $\phi(t)$ is any generic variable and $\Phi^{(h)}(t, \eta, \bar \eta)$ and $\Phi^{(d)}(t, \eta, \bar \eta)$ are 
the corresponding supervariables listed in (13), (19) and (34), respectively.

It is worthwhile to mention that the conserved  BRST and anti-BRST charges can be written in terms of the (anti-)BRST 
symmetry transformations as follows 
\begin{eqnarray}
Q_b &=& -\, i\,s_b\, \big(\bar C \, \dot C - \dot {\bar C}\, C\big) =  - \,i\, s_{ab} \,\big(\dot C \, C\big),  \nonumber\\
Q_{ab} &=& +\,i \,s_{ab}\, \big(\bar C \, \dot C - \dot {\bar C}\, C\big) =  +\,i\, s_b \,\big(\dot {\bar C} \, \bar C\big).
\end{eqnarray}
Exploiting the expressions of the supervariables given in (13) and (19), one can generalize these conserved charges onto 
$(1, 2)-$dimensional supermanifold as    
\begin{eqnarray}
Q_b &=& -\, i \, \frac{\partial}{\partial \bar \eta}\, \Big[\bar {\cal F}^{(h)}\, {\dot {\cal F}}^{(h)} 
- {\dot {\bar {\cal F}}}^{(h)} \, {\cal F}^{(h)} \Big] \bigg|_{\eta = 0} 
\equiv -\, i \int d\bar \eta \, \Big[\bar {\cal F}^{(h)}\, {\dot {\cal F}}^{(h)} 
- {\dot {\bar {\cal F}}}^{(h)} \, {\cal F}^{(h)} \Big] \bigg|_{\eta = 0}\nonumber\\
&=& -\, i\, \frac{\partial}{\partial \eta} \Big[\dot {\cal F}^{(h)}\, {\cal F}^{(h)}\Big]
\equiv -\, i \int d\eta \left[\dot {\cal F}^{(h)}\, {\cal F}^{(h)}\right],  \nonumber\\
Q_{ab} &=& +\, i \, \frac{\partial}{\partial \eta}\, \Big[\bar {\cal F}^{(h)}\, {\dot {\cal F}}^{(h)} 
- {\dot {\bar {\cal F}}}^{(h)} \, {\cal F}^{(h)}\Big] \bigg|_{\bar \eta = 0} 
\equiv +\, i \int d\eta\, \Big[\bar {\cal F}^{(h)}\, {\dot {\cal F}}^{(h)} 
- {\dot {\bar {\cal F}}}^{(h)} \, {\cal F}^{(h)}\Big] \bigg|_{\bar \eta = 0}  \nonumber\\
&=& +\, i\, \frac{\partial}{\partial \bar \eta} \Big[{\dot {\bar {\cal F}}}^{(h)}\, \bar {\cal F}^{(h)} \Big] 
\equiv +\, i \int  d \bar \eta\Big[{\dot {\bar {\cal F}}}^{(h)}\, \bar {\cal F}^{(h)}\Big]. 
\end{eqnarray}
Using the basic principles of BRST formalism, we can also write the BRST and anti-BRST charges in the following fashion; namely, 
\begin{eqnarray}
Q_b &=& i\, s_b\, s_{ab} \, \big(z\, C\big) = \frac{i}{2}\,s_b\, s_{ab}\, \big(\dot \theta \,C - \theta\, \dot C\big),  \nonumber\\   
Q_{ab} &=& i\, s_b\, s_{ab} \, \big(z\, \bar C\big) = \frac{i}{2}\,s_b\, s_{ab}\, \big(\dot \theta \,\bar C - \theta\, \dot {\bar C}\big).
\end{eqnarray}
From the expressions given in (50) and (52), it is quite easy to show that $s_b\,Q_b =0, \; s_{ab}\, Q_{ab} = 0$ 
which imply the  nilpotency properties:  $Q^2_b = 0, \;  Q^2_{ab} = 0$ whereas  $s_b\,Q_{ab} = 0,\; s_{ab}\, Q_b =0$ 
show the anticommutativity $Q_{ab}\, Q_b + Q_b\, Q_{ab} = 0$ of the  (anti-)BRST charges $Q_{(a)b}$.  
The above expressions for the (anti-)BRST charges  in terms of the supervariables are listed as follows
\begin{eqnarray}
Q_b &=& i \, \frac{\partial}{\partial \bar \eta}\, \frac{\partial}{\partial  \eta} \Big[{\cal Z}^{(h)}\, {\cal F}^{(h)}\Big]
\equiv i \, \int d\bar\eta \int d\eta \Big[{\cal Z}^{(h)}\, {\cal F}^{(h)}\Big]\nonumber\\
&=& \frac{i}{2}\,\frac{\partial}{\partial \bar \eta}\, \frac{\partial}{\partial  \eta} \Big[\dot \Theta^{(h)} \,{\cal F}^{(h)} 
- \Theta^{(h)}\, \dot {\cal F}^{(h)}\Big] 
\equiv \frac{i}{2}\,\int d\bar \eta \int d\eta \Big[\dot \Theta^{(h)} \,{\cal F}^{(h)} - \Theta^{(h)}\, \dot {\cal F}^{(h)}\Big], \nonumber\\
&& \nonumber\\
Q_{ab} &=& i \, \frac{\partial}{\partial \bar \eta}\, \frac{\partial}{\partial  \eta} \Big[{\cal Z}^{(h)}\, {\bar {\cal F}}^{(h)}\Big]
\equiv i  \int d\bar \eta  \int d\eta \Big[{\cal Z}^{(h)}\, {\bar {\cal F}}^{(h)}\Big]\nonumber\\
&=& \frac{i}{2}\,\frac{\partial}{\partial \bar \eta}\, \frac{\partial}{\partial  \eta} \Big[\dot \Theta^{(h)} \, {\bar {\cal F}}^{(h)} 
- \Theta^{(h)}\, {\dot {\bar {\cal F}}}^{(h)}\Big] 
\equiv \frac{i}{2} \int d\bar \eta \int  d\eta \Big[\dot \Theta^{(h)} \, {\bar {\cal F}}^{(h)} 
- \Theta^{(h)}\, {\dot {\bar {\cal F}}}^{(h)}\Big].
\end{eqnarray}
As a consequence of the expressions (51) and (53),  we can capture the nilpotency and anticommutativity of the (anti-)BRST charges 
in terms of the Grassmannian generators as given below:     
\begin{eqnarray}
&& \frac{\partial}{\partial \bar \eta}\,  Q_b = 0 \;\Leftrightarrow \; Q^2_b = 0, \qquad 
\frac{\partial}{\partial \eta}\,  Q_{ab} = 0 \;\Leftrightarrow \; Q^2_{ab} = 0,\nonumber\\
&&\nonumber\\
&& \frac{\partial}{\partial \bar \eta}\,  Q_{ab} = \frac{\partial}{\partial \eta}\,  Q_b = 0 \;\Leftrightarrow \; Q_b\, Q_{ab} + Q_{ab}\, Q_b = 0.
\end{eqnarray} 
This algebra is true because of the the fact that $\partial^2_\eta = 0,\; \partial^2_{\bar \eta} = 0$ and 
$\partial_\eta\, \partial_{\bar \eta} + \partial_{\bar \eta }\, \partial_\eta = 0$.

In a similar fashion, we can write the co-BRST and anti-co-BRST charges, in four different ways; namely,
\begin{eqnarray}
Q_d &=& i\,s_d\, \big(\bar C \, \dot C - \dot {\bar C}\, C\big) = i\, s_{ad} \,\big(\dot {\bar C} \, \bar C\big) \nonumber\\               
    &=& i\, s_d\, s_{ad} \, \big(\theta \,\bar  C\big) = \frac{i}{2}\,s_d\, s_{ad}\, \big(z \, \dot{\bar C} - \dot z \,  {\bar C}\big), \nonumber\\
Q_{ad} &=& - \,i \,s_{ad}\, \big(\bar C \, \dot C - \dot {\bar C}\, C\big) =  - \, i\, s_d \,\big(\dot C \, C\big)\nonumber\\   
&=& i\, s_d\, s_{ad} \, \big(\theta \,  C\big) = \frac{i}{2}\,s_d\, s_{ad}\, \big(z \, \dot C - \dot z \, C\big).
\end{eqnarray}
It is clear from the above expressions for the conserved (i.e. $\dot Q_{(a)d} = 0$) (anti-)co-BRST charges $Q_{(a)d}$, 
one can now again easily show $Q^2_d = 0, \;Q^2_{ad} = 0$ and 
$Q_d\, Q_{ad} + Q_{ad}\, Q_d = 0$ by exploiting the definition of a generator. For instance, the following relation $s_d Q_d = - i \{Q_d, \, Q_d\} = 0$ 
leads to $Q^2_d = 0$ which shows the nilpotency of co-BRST charge.

In terms of the supervariables (34), the (anti-)co-BRST charges given in (55) take the following forms:  
\begin{eqnarray}
Q_d &=& +\, i \, \frac{\partial}{\partial \bar \eta}\, \Big[\bar {\cal F}^{(d)}\, {\dot {\cal F}}^{(d)} 
- {\dot {\bar {\cal F}}}^{(d)} \, {\cal F}^{(d)} \Big] \bigg|_{\eta = 0} 
\equiv +\, i \int d\bar \eta \, \Big[\bar {\cal F}^{(d)}\, {\dot {\cal F}}^{(d)} 
- {\dot {\bar {\cal F}}}^{(d)} \, {\cal F}^{(d)} \Big]\bigg|_{\eta = 0}\nonumber\\
&=& +\, i\, \frac{\partial}{\partial \eta} \Big[{\dot {\bar {\cal F}}}^{(d)}\,{\bar  {\cal F}}^{(d)}\Big] 
\equiv +\, i \int d\eta \Big[{\dot {\bar {\cal F}}}^{(d)}\,{\bar  {\cal F}}^{(d)}\Big] \nonumber\\
&=& i \, \frac{\partial}{\partial \bar \eta}\, \frac{\partial}{\partial  \eta} \Big[\Theta^{(d)}\, {\bar {\cal F}}^{(d)}\Big]
\equiv i \, \int d\bar\eta \int d\eta \Big[\Theta^{(d)}\, {\bar {\cal F}}^{(d)}\Big]\nonumber\\
&=& \frac{i}{2}\,\frac{\partial}{\partial \bar \eta}\, \frac{\partial}{\partial  \eta} \Big[{\cal Z}^{(d)} \,{\dot {\bar {\cal F}}}^{(d)} 
- \dot {\cal Z}^{(d)}\, {\bar  {\cal F}}^{(d)}\Big] 
\equiv \frac{i}{2}\,\int d\bar \eta \int d\eta \Big[{\cal Z}^{(d)} \,{\dot {\bar {\cal F}}}^{(d)} 
- \dot {\cal Z}^{(d)}\, {\bar  {\cal F}}^{(d)}\Big], \nonumber\\
&&\nonumber\\
Q_{ad} &=& -\, i \, \frac{\partial}{\partial \eta}\, \Big[\bar {\cal F}^{(d)}\, {\dot {\cal F}}^{(d)} 
- {\dot {\bar {\cal F}}}^{(d)} \, {\cal F}^{(d)}\Big] \bigg|_{\bar \eta = 0} 
\equiv - i \int d\eta\, \Big[\bar {\cal F}^{(d)}\, {\dot {\cal F}}^{(d)} 
- {\dot {\bar {\cal F}}}^{(d)} \, {\cal F}^{(d)}\Big] \bigg|_{\bar \eta = 0}   \nonumber\\
&=& -\, i\, \frac{\partial}{\partial \bar \eta} \Big[\dot {\cal F}^{(d)}\,  {\cal F}^{(d)} \Big]
\equiv -\, i \int  d \bar \eta\Big[\dot {\cal F}^{(d)}\,  {\cal F}^{(d)}\Big] \nonumber\\
&=& i \, \frac{\partial}{\partial \bar \eta}\, \frac{\partial}{\partial  \eta} \Big[\Theta^{(d)}\, {\cal F}^{(d)}\Big]
\equiv i \, \int d\bar\eta \int d\eta \Big[\Theta^{(d)}\, {\cal F}^{(d)}\Big]\nonumber\\
&=& \frac{i}{2}\,\frac{\partial}{\partial \bar \eta}\, \frac{\partial}{\partial  \eta} \Big[{\cal Z}^{(d)} \,{\dot {\cal F}}^{(d)} 
- \dot {\cal Z}^{(d)}\,  {\cal F}^{(d)}\Big] 
\equiv \frac{i}{2}\,\int d\bar \eta \int d\eta \Big[{\cal Z}^{(d)} \,{\dot{\cal F}}^{(d)} 
- \dot {\cal Z}^{(d)}\,  {\cal F}^{(d)}\Big].
\end{eqnarray}
It clear from the above equation that the following relations are true; namely,
\begin{eqnarray}
&&\frac{\partial}{\partial \bar \eta}\,  Q_d = 0 \,\Leftrightarrow \, Q^2_d = 0, \quad 
\frac{\partial}{\partial \eta}\,  Q_{ad} = 0 \,\Leftrightarrow \, Q^2_{ad} = 0,\nonumber\\
&& \frac{\partial}{\partial \bar \eta}\,  Q_{ad} = \frac{\partial}{\partial \eta}\,  Q_d = 0 \;\Leftrightarrow \; Q_d\, Q_{ad} + Q_{ad}\, Q_d = 0,
\end{eqnarray} 
where we have used the properties of the translational generators $\partial_\eta$ and 
$\partial_{\bar \eta}$. It is now clear that we have captured  the nilpotency as well as anticommutativity properties of the fermionic 
symmetry transformations (and corresponding charges) in the language of translational generators along the Grassmannian directions $(\eta, \bar \eta)$.


\section{Conclusions}


In our present endeavor, we have derived the {\it proper} off-shell nilpotent and  absolutely anticommuting (anti-)BRST as well as (anti-)co-BRST
symmetry transformations within the framework of ``augmented" supervariable approach. For the derivation of (anti-)BRST symmetry transformations, 
we have used, on one hand, horizontality condition and gauge-invariant restriction [cf. (10) and (17)]. On the other hand, we have exploited 
the dual-horizontality condition together with (anti-)co-BRST invariant restrictions for the precise derivation of (anti-)co-BRST  
transformations [cf. (24), (30) and (32)]. The (anti-)BRST and (anti-)co-BRST transformations for the Nakanishi--Lautrup type variable $b$ have been derived 
from the requirements of the nilpotency and absolute anticommutativity properties of these transformations.

We point out that the first-order Lagrangian $L_f$ is gauge-invariant and (anti-)BRST invariant. As a consequence,
$L_f$ is independent of the Grassmannian variables 
[cf. (38)] when it is generalized onto $(1, 2)$D superspace. 
Further, the total gauge--fixing terms remain invariant under the (anti-)co-BRST symmetry transformations. In the superspace, one can see from (46) 
that it is also independent of Grassmannian variables $(\eta, \bar \eta)$.  We have expressed the total Lagrangian (3)
in terms of the continuous and nilpotent symmetry transformations $s_{(a)b}$ and $s_{(a)d}$.  In fact,
for the (anti-)BRST invariance, the total gauge--fixing and Faddeev--Popov ghost terms can be written as the BRST-exact, anti-BRST exact and
anti-BRST exact followed by BRST exact [cf. (21)]. Similarly, for the (anti-)co-BRST invariance, the total Lagrangian (3) is also written in terms of the 
(anti-)co-BRST symmetry transformations [cf. (36)]. Thus, the (anti-)BRST and (anti-)co-BRST invariances of the Lagrangian (3) become straightforward
because of the nilpotency property of the symmetry transformations $s_{(a)b}$ and $s_{(a)d}$.

We have provided the geometrical origin of the continuous (anti-)BRST and (anti-)co-BRST transformations within the framework of superspace formalism
[cf. (20) and (35)].  By using the basic tenets of supervariable approach, we have written the Lagrangian in many different ways in terms of the 
supervariables (13) and (19) for the (anti-)BRST invariance and in terms of supervariables (34) for the (anti-)co-BRST invariance.
 Thus,  we have been able to capture 
the invariance of the Lagrangian in the language of translational generators $\partial_\eta$ and $\partial_{\bar \eta}$ (cf., Section 5).

Further, we have expressed the (anti-)BRST and (anti-)co-BRST charges in terms of  the nilpotent symmetry transformations 
[cf. (50), (52) and (55), respectively].  In view of these, it is easy for us to write the conserved charges in terms of the 
supervariables, as one can see, in equations (51), (53) and (56). The key properties (i.e. nilpotency and anticommutativity) 
associated with the (anti-)BRST and (anti-)co-BRST transformations (and corresponding conserved charges) are translated in the 
properties of Grassmannian translational generators $\partial_\eta, \; \partial_{\bar \eta}$ along $\eta,\; \bar \eta$ directions.

\section*{Acknowledgments}

RK would like to thank UGC, Government of India, New Delhi, for financial support under the PDFSS scheme. 
AS would like to thank Prof. Shu Lin for his kind support.

\end{document}